# The Stellar Imager (SI) Vision Mission

Kenneth G. Carpenter[a*], Carolus J. Schrijver[b], Margarita Karovska[c],
and the SI Vision Mission Study Team

[a]NASA-GSFC, Code 667, Greenbelt, MD 20771
[b]Lockheed Martin Advanced Technology Center, Palo Alto, CA 94304
[c]Smithsonian Astrophysical Observatory, Cambridge, MA 02138

## ABSTRACT

The Stellar Imager (SI) is a UV-Optical, Space-Based Interferometer designed to enable 0.1 milli-arcsecond (mas) spectral imaging of stellar surfaces and of the Universe in general and asteroseismic imaging of stellar interiors. SI is identified as a "Flagship and Landmark Discovery Mission" in the 2005 Sun Solar System Connection (SSSC) Roadmap and as a candidate for a "Pathways to Life Observatory" in the Exploration of the Universe Division (EUD) Roadmap (May, 2005). SI will revolutionize our view of many dynamic astrophysical processes: its resolution will transform point sources into extended sources, and snapshots into evolving views. SI's science focuses on the role of magnetism in the Universe, particularly on magnetic activity on the surfaces of stars like the Sun. SI's prime goal is to enable long-term forecasting of solar activity and the space weather that it drives. SI will also revolutionize our understanding of the formation of planetary systems, of the habitability and climatology of distant planets, and of many magneto-hydrodynamically controlled processes in the Universe. The results of the SI "Vision Mission" Study are presented in this paper. Additional information on the SI mission concept and related technology development can be found at URL: http://hires.gsfc.nasa.gov/si/.

## 1. INTRODUCTION

The Stellar Imager (SI) is a UV-Optical, Space-Based Interferometer designed to enable 0.1 milli-arcsecond (mas) spectral imaging of stellar surfaces and stellar interiors (via asteroseismology) and of the Universe in general. At the revolutionary design resolution of SI, sequences of images will show the dynamics of astrophysical processes and perhaps even allow us to directly see, for the first time, the evolution of, e.g., a planetary nebula, an early supernova phase, mass exchange in binaries, (proto-)stellar jets, and/or accretion systems. Its spectral imaging capability is designed to enable an improved understanding of:

- o **Solar and Stellar Magnetic Activity and Its Impact on Space Weather, Planetary Climates, and Life**
- o **Magnetic Processes, the Origin and Evolution of Structure, and the Transport of Matter Throughout the Universe**

SI is included as a "Flagship and Landmark Discovery Mission" in the 2005 Sun Solar System Connection (SSSC) Roadmap and as a candidate for a "Pathways to Life Observatory" in the Exploration of the Universe Division (EUD) Roadmap (May, 2005).

This paper summarizes the Final Report of the SI Vision Mission (VM) Study, for which the principal authors and their primary areas of expertise are listed in the full VM Report. A one-page "Quick Facts" sheet summarizing the Mission's Goals and Architecture is given in **Table 1.** The SI concept, initiated by Lockheed Martin, is under development by NASA's Goddard Space Flight Center, in collaboration with a broad variety of industrial, academic, and astronomical science institute partners, as well as an international group of science and technical advisors.

Further information on SI, its science, and the technology development needed to enable its realization can be found at *http://hires.gsfc.nasa.gov/si/*.

*Kenneth.G.Carpenter@nasa.gov; phone 1 301-286-3453; fax 1 301 286 1753

Table 1: Quick Facts: The Stellar Imager (SI) Vision Mission

| Mission Overview | | |
|---|---|---|
| *SI is a UV-Optical, Space-Based Interferometer for 0.1 milli-arcsecond (mas) spectral imaging of stellar surfaces and stellar interiors (via asteroseismology) and of the Universe in general.* | | |
| **Science Goals** | | |
| To understand:<br>- Solar and Stellar Magnetic Activity<br>    and their impact on Space Weather, Planetary Climates, and Life<br>- Magnetic Processes and their roles in the Origin and Evolution of Structure<br>    and in the Transport of Matter throughout the Universe | | |
| **Mission and Performance Parameters** | | |
| **Parameter** | **Value** | **Notes** |
| Maximum Baseline (B) | 100 – 1000 m (500 m typical) | Outer array diameter |
| Effective Focal Length | 1 – 10 km     (5 km typical) | Scales linearly with B |
| Diameter of Mirrors | 1 - 2 m       (1 m currently) | Up to 30 mirrors total |
| λ-Coverage | UV:     1200 – 3200 Å<br>Optical: 3200 – 5000 Å | Wavefront Sensing in optical only |
| Spectral Resolution | UV: 10 Å (emission lines)<br>UV/Opt: 100 Å (continuum) | |
| Operational Orbit | Sun-Earth L2 Lissajous, 180 d | 200,000x800,000 km |
| Operational Lifetime | 5 yrs (req.) – 10 yrs (goal) | |
| Accessible Sky | Sun angle: 70º ≤ β ≤ 110º | Entire sky in 180 d |
| Hub Dry Mass | 1455 kg | Possibly 2 copies |
| Mirrorsat Dry Mass | 65 kg (BATC) - 120 kg (IMDC) | For each of up to 30 |
| Ref. Platform Mass | 200 kg | |
| Total Propellant Mass | 750 kg | For operational phase |
| Angular Resolution | 50 µas – 208 µas (@1200–5000Å) | Scales linearly ~ λ/B |
| Typical total time to image stellar surface | < 5 hours for solar type<br>< 1 day for supergiant | |
| Imaging time resolution | 10 – 30 min (10 min typical) | Surface imaging |
| Seismology time res. | 1 min cadence | Internal structure |
| # res. pixels on star | ~1000 total over disk | Solar type at 4 pc |
| Minimum FOV | > 4 mas | |
| Minimum flux detectable at 1550 Å | $5.0 \times 10^{-14}$ ergs/cm$^2$/s integrated over C IV lines | 10 Å bandpass |
| Precision Formation Fly. | s/c control to mm-cm level | |
| Optical Surfaces Control | Actuated mirrors to µm-nm level | |
| Phase Corrections | to λ/10 Optical Path Difference | |
| Aspect Control/Correct. | 3 µas for up to 1000 sec | Line of sight mainten. |

## 2. SCIENCE RATIONALE

### 2.1 Key objectives

The key science goals of the SI mission are:

- **To study the evolution of stellar magnetic dynamos** from the very formation of stars and planetary systems onward to the final stages of stellar evolution.

- **To complete the assessment of external solar systems** begun with the planet-finding and imaging missions **by imaging their central stars**

- **To study the Universe at ultra-high angular resolution** from the internal structure and dynamics of stars and interacting binaries to extreme conditions, e.g. in Active Galactic Nuclei and black hole environments.

### 2.2 Primary science goals for the Stellar Imager: stellar magnetic activity

Most of us rarely give the Sun a second thought. We do not question its presence or its apparent stability as we see it traverse the sky every day. The Sun is, however, *a variable* star. Its variability affects the Earth and the human society by modulating Earth's climate. It also affects our technology, upon which we are becoming ever more reliant: eruptions on the Sun disrupt communications; affect navigation systems; cause radiation harmful to astronauts exploring beyond the Earth's atmosphere and to airline passengers traveling through it; and occasionally push power grids to fail.

The recognition of the importance of the Sun's fickle variations has led to the development of a large National Space Weather Architecture. Within that Architecture, NASA, and in particular the Heliophysics Division is working to learn why and how Earth and human society are affected by the Sun's variable magnetism. This is the focus of NASA's *Living With A Star* program. At the core of that program is the question concerning the Sun's magnetic field: what causes the Sun to be magnetically active, and how can we develop reliable forecasting tools for this activity and the associated space weather and climate changes on Earth? The Stellar Imager aims to make crucial contributions to this field, warranting its status as a Landmark Discovery Mission in the 2005 roadmap for the Sun-Earth Connection.

The principal cause of all solar variability is its magnetic field. This intangible and unfamiliar fundamental force of nature is created in the convective envelope of the Sun by a process that we call the dynamo. There is at present no quantitative model for stellar dynamos that is useful to forecast solar activity or even to establish the mean activity level of a star based on, say, its mass, age, and rotation rate. The nonlinear differential equations for the coupling of the vectors of turbulent convection and magnetic field cannot be solved analytically. Nor can the cycle dynamo be simulated numerically in its entirety; full numerical coverage would require some $10^{18}$ grid points, which is a factor of order a billion beyond present computational means. Hence, both analytical and numerical studies necessarily make approximations that simplify or ignore much of the physics. Furthermore, even the approximating models are of a richness and diversity that there is no consensus on the model properties, or even on the set of processes that are important in driving the dynamo. Numerical research will undoubtedly make significant advances in the coming years, but only the comparative analysis of many Sun-like stars with a range of activity levels, masses, and evolutionary stages will allow adequate tests of complex dynamo models, validation of any detailed dynamo model, and exploration of the possible spatio-temporal patterns of the nonlinear dynamo.

The studies of average activity levels of stars have helped us piece together what some of the essential ingredients to dynamo action are on the largest scales. For example, we know that a dynamo associated with stellar activity operates in all rotating stars with a convection zone directly beneath the photosphere. In single stars, the dynamo strength varies smoothly, and mostly monotonically, with rotation rate, at least down to the intrinsic scatter associated with stellar variability. It also depends on some other unknown stellar property or properties. For main sequence stars, for example, the primary factor in determining activity resembles the convective turnover time scale at the bottom of the convective envelope. But no such dependence holds if we test the relationship on either evolved stars or on tidally-interacting compact binary systems. Apparently, other parameters, as yet unidentified, play a role, such as surface gravity and tidal forces.

The variations of stellar and solar activity on time scales of years also remain a mystery. The Sun shows a relatively regular heartbeat with its 11-year sunspot cycle, even as cycle strength and duration are modulated. Such a pattern is not the rule among the cool main–sequence stars, however. Instead, we find a variety of patterns of variability in their activity, in which *only one in three* of these stars show cyclic variations that resemble those of the Sun. For main–sequence stars with moderate to low rotation rates, activity tends to be cyclic, but no clear trend of cycle period with stellar parameters has been found, although there are hints of relationships between cycle period, rotation period, and the time scale for deep convection. For truly active stars, various variability patterns exist, but generally no unambiguous activity cycle is seen.

Historical records show that the Sun can change its activity significantly on the intermediate time scale of decades (**see Fig. 1**). Activity decreased, for example, for multiple decades during the 17th Century, when Earth experienced the Little Ice Age. A sustained increase in activity – such as happened during the medieval Grand Maximum – may cause a warm spell, and will be associated with an increase in the frequency of space storms, and in the ultraviolet radiation that is harmful to life on Earth.

*It would take hundreds of years to validate a solar dynamo model using only observations of the Sun, given its **irregular** 11-year magnetic heartbeat and the long-term modulations.* Key to successfully navigating the route to a workable, predictive dynamo model is the realization that *in order to understand the solar dynamo, we need a population study;*

*that is, we need to study the dynamo-driven activity in a sample of stars like the Sun, and compare it to observations of younger stars, older stars, and stars in binary systems, etc*. Thus, the SI will enable us to test and validate solar dynamo models *within a decade, rather than requiring a century or more* if we used only the Sun.

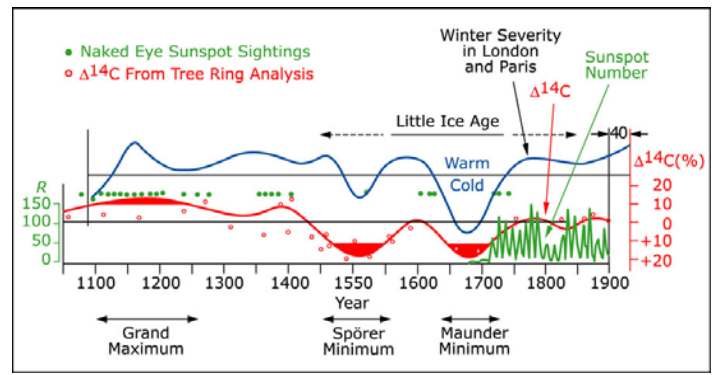

Fig. 1. Short term variations in Solar activity and their impact on Earth.

The potential for a breakthrough in our understanding and our prediction ability lies in spatially-resolved imaging of the dynamo-driven activity patterns on a variety of stars. These patterns, and how they depend on stellar properties (including convection, differential rotation and meridional circulation, evolutionary stage/age), are crucial for dynamo theorists to explore the sensitive dependences on many poorly known parameters, to investigate bifurcations in a nonlinear 3-dimensional dynamo theory, and to validate the ultimate model.

*Direct, interferometric imaging – the goal of the Stellar Imager - is the only way to obtain the required information on the dynamo patterns for stars of Sun-like activity.* Alternative methods that offer limited information on spatial patterns on much more active stars fail for a Sun-like star: a) rotationally-induced Doppler shifts in such stars are too small compared to the line width to allow Zeeman-Doppler imaging, b) the activity level is insufficient to lead to significant spectral changes associated with magnetic line splitting, c) rotational modulation measurements leave substantial ambiguities in the latitude distributions, locations and sizes of spots, and cannot be used to measure dispersal of field across the stellar surface. The direct imaging by SI of stellar activity will overcome these problems. Equally importantly, the asteroseismic observations planned with SI will determine the internal properties of stellar structure and rotation, thus directly providing crucial information relevant to the physical operation of the dynamo mechanism.

Imaging magnetically active stars and their surroundings will also provide us with an indirect view of the Sun through time, from its formation in a molecular cloud, through its phase of decaying activity, during and beyond the red-giant phase during which the Sun will swell to about the size of the Earth's orbit, and then toward the final stages of its evolution as a Planetary Nebula and a white dwarf relic.

The SI mission will allow us not only to image the surfaces of stars, but also to sound stellar interiors using spatially resolved asteroseismology to image internal structure, differential rotation, and large-scale circulations; this will provide accurate knowledge of stellar structure and evolution and complex transport processes, and will impact numerous branches of (astro)physics. *For arrays of 9 or more optical elements, asteroseismic imaging of structure and rotation is possible with a depth resolution of 20,000 km for a star like the Sun.*

Helioseismology has given us an extremely detailed view of the solar interior. These results are of great importance to our understanding of the structure and evolution of stars, and of the physical properties and processes that control this evolution. At the time of the launch of the SI, seismic investigations of other stars will have been undertaken by several space missions, including MOST and COROT, however, a number of key issues will remain open. These missions will only observe low-degree modes, through intensity variations in light integrated over the stellar disks. Such point-source observations will provide information about the global properties of solar-like stars, which allows the study of global structure, including, e.g., gravitational settling of helium and large-scale mixing processes. SI observations, however, will allow us to expand the discovery space far beyond that: modes of degree as high as 60 should be reachable with an array of N=10 elements, increasing as $N^2$ for larger arrays. By analogy with the Sun, in solar-like stars this will allow inferences with good radial and reasonable latitude resolution to be made in the radiative interior and the lower part of the convective envelope, for both structure and the patterns and magnitudes of the differential rotation with depth and latitude. With a careful choice of target stars SI observations will allow us to obtain such detailed information about the interiors of stars over a broad range of stellar parameters, in terms of mass, age and composition.

Studies of the internal rotation as a function of mass and age will provide unique information about the evolution of stellar internal rotation with age, in response to the activity-driven angular-momentum loss in stellar winds. This will provide stringent constraints on models of the rotational evolution, elucidating the processes responsible for transport of angular momentum in stellar interiors; these studies are also fundamental to the understanding of the dynamo processes likely responsible for stellar activity. By correlating the rotation profile with the profile of the helium abundance, as reflected in the seismically inferred sound speed, an understanding can be achieved of the rotationally-driven mixing processes in stellar interiors. This is of great importance for calibrating the primordial abundances in the Universe as well as to the improvement and validation of stellar evolution models. For example, the data will provide constraints on the convective overshoot at the base of the convective envelope which also contributes to the mixing. The resulting understanding can then be applied to the mixing and destruction of lithium, finally providing the means to relate the observed lithium abundance in old halo stars to the primordial lithium content of the Universe. For stars slightly more massive than the Sun the data, combined with the more extensive data on low-degree modes likely available at the time from earlier missions, will allow detailed investigations of the properties of convective cores and related internal mixing; an understanding of these processes is essential to the modeling of the evolution of massive stars, leading to the formation of supernovae.

In summary, *the UV/optical imaging of stellar surfaces, combined with the asteroseismic study of their interiors, at high spatial and temporal resolution will enable a tremendous leap forward in our understanding of stellar magnetic activity and its impact on stellar structure and evolution and, most importantly, on the climate of surrounding planets and on the ability of those planets to originate and maintain life.*

### 2.3 The Universe at ultra-high angular resolution

A long-baseline interferometer in space will benefit many fields of astrophysics and physics. In particular, magnetic fields affect the evolution of stars and planetary systems in all phases, from the formation of the star and its planets, to the habitability of these planets through the billions of years during which they live with their stars. With its revolutionary imaging power, *SI will enable detailed study of magnetic processes and their roles in the Origin and Evolution of Structure and in the Transport of Matter throughout the Universe*.

Examples of additional scientific areas of study for the Stellar Imager include:

- **Stellar interiors** *in stars outside solar parameters*
- **Infant Stars-disk systems** *to image dynamic accretion, magnetic field structure & star/disk interaction*
- **Hot Stars** *and their hot polar winds, non-radial pulsations, rotation, structure, and the envelopes and shells of Be-stars*
- **Cool, Evolved Giant & Supergiant Stars** *and the spatiotemporal structure of extended atmospheres, pulsation, winds, shocks*
- **Supernovae & Planetary Nebulae:** *their core structure, early expansion and interaction with CSM*
- **Interacting Binaries,** *including mass-exchange, dynamical evolution, accretion, and dynamos*
- **Active Galactic Nuclei, Quasars, Black-Hole Environments**, etc. …

SI will produce images with hundreds of times more detail than Hubble. **Figure 2** shows examples of SI snapshot views of a solar-type star at 4 pc, and of diverse galactic and extragalactic sources that are far beyond the reach of the current and near future observational astronomy. Furthermore, the SI will bring the study of dynamical evolution of many astrophysical objects into reach: hours to weeks between successive images will detect dramatic changes in many objects, e.g., mass transfer in binaries, pulsation-driven surface brightness variation and convective cell structure in giants and supergiants, jet formation and propagation in young planetary systems, reverberating Active Galactic Nuclei (AGN), and many others. Imagine, for example, unprecedented dynamic views of evolving structures of AGN, quasi-stellar objects, supernovae, interacting binary stars, supergiant stars, hot main-sequence stars, star-forming regions, and protoplanetary disks. Additional details are given the full VM report.

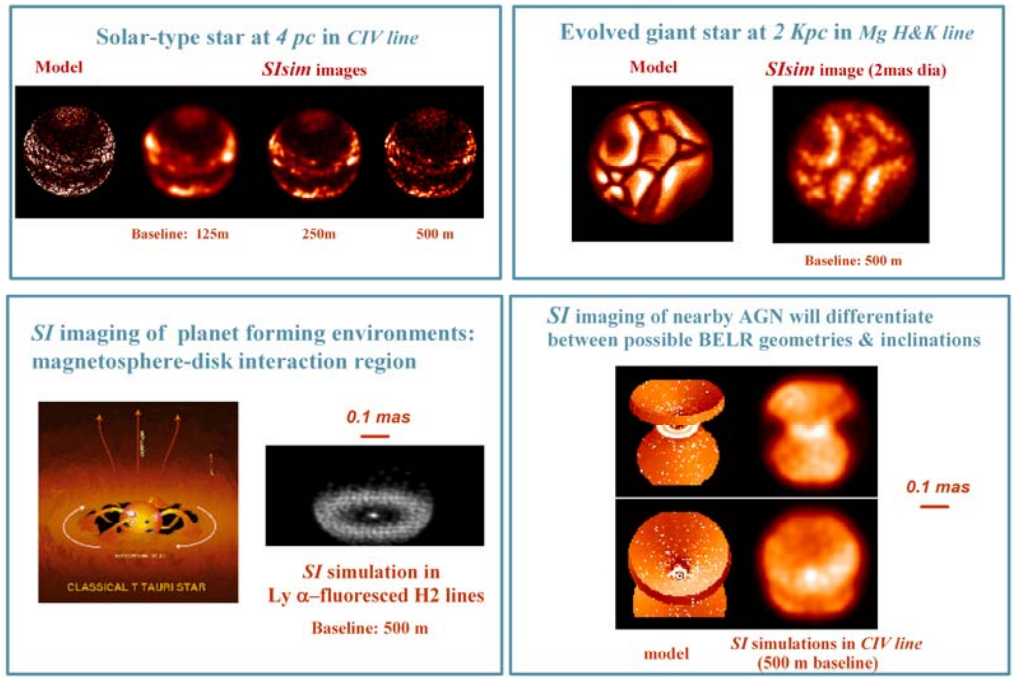

Fig. 2. Simulations of some of SI's capabilities for UV imaging, assuming 30 mirror elements in a non-redundant pattern.

**2.4 Relation to NASA and SMD Strategic Plans and Other Projects**

Fitting naturally within the NASA long-term time line, SI complements defined and proposed missions (Terrestrial Planet Finder – I, Life Finder, and Planet Imager), and with them will show us entire other solar systems, from the central star to their orbiting planets. It moreover fits on the technology roadmap that leads from interferometers like Keck and SIM to TPF-I/Darwin, MAXIM/Black Hole Imager, Life Finder, and the Planet Imager.

Stellar Imager was included in the 2000 and 2003 SEC Roadmaps and is now identified as a "*Flagship and Landmark-Discovery Mission*" in the 2005 Sun Solar System Connection (SSSC) Roadmap. SI is also a candidate for a "*Pathways to Life Observatory*" in the Exploration of the Universe Division (EUD) Roadmap (May, 2005). SI will provide an angular resolution over 200x that of the Hubble Space Telescope (HST) and will resolve for the first time the surfaces of Sun-like stars and the details of many astrophysical objects and processes. Stellar Imager is a natural culmination of science addressed with ongoing ground-based observatories and a series of space missions. These efforts will provide information on long-term disk-integrated variability, large-scale internal structure and evolutionary status, distances and other fundamental stellar properties, binary properties, and low-resolution surface imaging for a subset of target classes. SI complements and builds on observations made by ground-based interferometers, by asteroseismology missions, JWST, and other missions. It complements the planet-finding missions by providing a view of the space-weather environment of the planetary systems studied in those missions, and thus provides critical data needed to understand fully which of the detected planets are indeed habitable.

The Stellar Imager fits in the national science priorities, the NASA strategic plan, the Living With A Star initiative, and the technology roadmap and meets scientific priorities identified by the National Academy of Sciences Astronomy and Astrophysics Survey Committee (2001, Ref. 1). With SI we can "survey the Universe and its constituents," "use the Universe as a unique laboratory," "study the formation of stars and their planetary systems, and the birth and evolution of giant and terrestrial planets," and, by focusing on the driver of space weather in past, present, and future, "understand how the astronomical environment affects Earth." *SI is responsive to a key national priority: imaging of magnetically active stars provides the only means to test any theory of solar magnetic activity as the driver of space weather and climate that can be achieved within a decade after launch.*

## 3. ARCHITECTURAL CONCEPT

The baseline full-mission concept for SI was developed in collaboration with the GSFC Integrated Mission Design Center (IMDC) and Instrument Synthesis and Analysis Lab (ISAL). In addition to assisting in the development of the architecture, the Design Centers explored the technical feasibility of the mission and identified the technology developments needed to enable the mission in the 2025 timeframe. The results of these IMDC and ISAL studies have been combined with related work carried out by the SI Vision Mission Study Team members to define the SI Concept.

The current baseline architecture concept (**Fig. 3**) for the full Stellar Imager (SI) mission is a space-based, UV-Optical Fizeau Interferometer with 20-30 one-meter primary mirrors, mounted on formation-flying "mirrorsats" distributed over a parabolic virtual surface whose diameter can be varied from 100 m up to as much as 1000 m, depending on the angular size of the target to be observed. The individual mirrors are fabricated as ultra-smooth, UV-quality flats and are actuated to produce the extremely gentle curvature needed to focus light on the beam-combining hub that is located at the prime focus from 1 – 10 km distant. The focal length scales linearly with the diameter of the primary array: a 100 m diameter array corresponds to a focal length of 1 km and a 1000 m array with a focal length of 10 km. The typical configuration has a 500 m array diameter and 5 km focal length. A one-meter primary mirror size was chosen to ensure that the primary stellar activity targets can be well observed with good signal/noise. Sizes up to two meters may be considered in the future, depending on the breadth of science targets that SI is required to observe – e.g., some fainter extragalactic objects may need larger mirrors, but those will come at a cost to the packaging for launch, the number of launches needed, and total mission cost. The mirrorsats fly in formation with a beam-combining hub in a Lissajous orbit around the Sun-Earth L2 point. The satellites are controlled to mm-micron radial precision relative to the hub and the mirror surfaces to 5 nm radial precision, rather than using optical delay lines inside the hub for fine tuning the optical path lengths. A second hub is strongly recommended to provide critical-path redundancy and major observing efficiency enhancements. The observatory may also include a "reference craft" to perform metrology on the formation, depending on which metrology design option is chosen (see full report for more details). **Fig. 3 (right)** provides an overview of the selected architecture: the upper panel shows a cross-sectional schematic of the entire observatory, while the lower panel shows a close-up of the hub and its major components.

The design and implementation plan for the SI does not require major improvements in infrastructure for a 2025 launch. Heavy lift vehicles in the Delta IV Heavy (or the future Atlas V heavy) are assumed available to launch the entire constellation in one or two launches – which are the most efficient ways to launch and deploy the observatory, though more numerous launches on smaller ELV's could be utilized if needed. Capabilities for supporting significant science and operations telecom data rates to/from Sun-Earth L2 are assumed (rough assumptions for SI data collection rates include 900 kbps daily average for 11 months/year and 5 Mbps average for 1 month/year). The most important capability not currently available would be the ability to reach and service facilities in Lissajous orbits around the L2 point. The long lifetime goal for SI suggests that it could benefit greatly from a human and/or robotic capability to refuel at a minimum and, optimally, service the various components of the mirrorsats and hub – and the design of all the spacecraft is envisioned as modular to enable servicing/exchange of the various important components.

Although the SI baseline design does not require that humans and/or robots be able to access and work on SI at the Sun-Earth L2 site, the mission could benefit greatly from such a capability. In particular, the long lifetime requirement for SI (5-10 years or more) is most easily met if the design can be made modular so that humans and/or robots can readily service and replace key components of the mirrorsats and hub. An obvious and simple capability that would help enable SI would be the ability to refuel the spacecraft to ensure it will be able to perform station-keeping/orbit maintenance and target-to-target maneuvering over the desired lifetime. Servicing of the critical hub spacecraft would also be of great utility, since it, unlike the mirrorsats, is a single-point failure, unless more than one hub is launched (or is available for launch-on-need). In-space servicing of the SI components will require provisions for access, capture, and handling by the servicing system visiting vehicles and robots or EVA astronauts. Standard features and modular designs greatly reduce the mission risks, costs, and operations impacts associated with servicing compared and are utilized in the SI design.

A timeline for the development of the SI mission or an equivalent long-baseline, UV/Optical, space-based interferometer depends on many unknowns at this early stage, but our study indicates that the needed technology developments described in the next section could be resolved in time to support a Pathfinder Mission in the 2015 timeframe and the launch of the full mission ~2025.

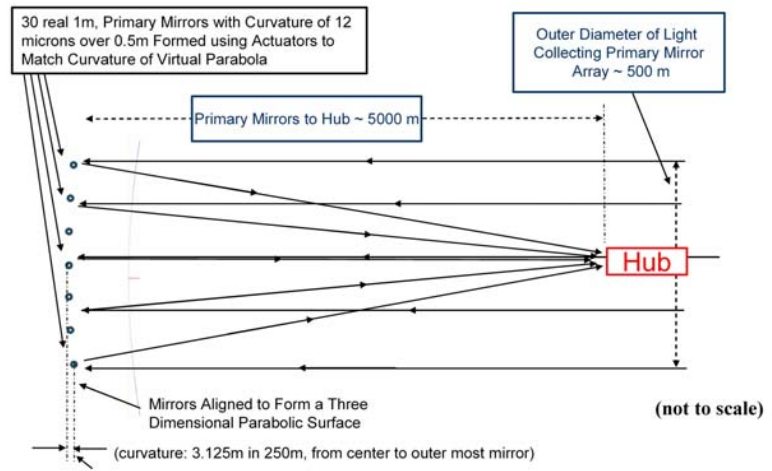

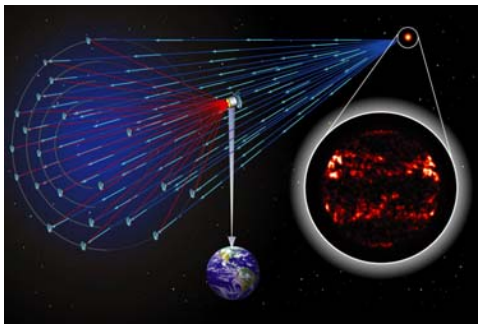

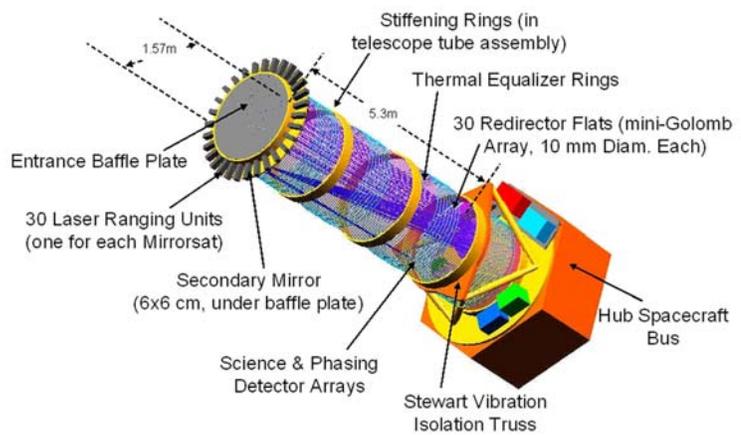

Fig. 3. The baseline SI design -  Left:  An artist's concept;  Right:  Schematics – cross-sectional view (top), hub (bottom)

## 4. TECHNOLOGY

### 4.1 Key enabling technologies, risks and uncertainties

The major technologies required to enable the SI mission as described in this paper are summarized in **Table 2**. Probably the most difficult of these is the precision formation flying of as many as 33 distinct spacecraft: 30 mirrorsats, 1-2 beam-combining hubs, and possibly a reference spacecraft for metrology and aspect control.  This is a complicated, multi-stage controls problem.  However, similar control systems will be needed for many future missions (e.g. at some level, all missions composed of distributed spacecraft flying in a formation with tight constraints), so there is a great deal of motivation for such development.  The biggest risk at the moment is the lack of a well-defined sequence of intermediate demonstration missions – with the cancellation of STARLIGHT, only SMART-3 and, possibly, ST9, are currently under consideration for flight prior to attempts at flying the large strategic missions like TPF-I, SI, LF, etc.  We propose to develop a Pathfinder mission to fill in this development "hole" and to prove other technologies (e.g., UV beam-combination) and pursue intermediate science goals as well – but more can and should be done.

Table 2: The major enabling technologies needed for Stellar Imager

- **formation-flying of ~30 spacecraft**
    - deployment and initial positioning of elements in large formations
    - real-time correction and control of formation elements
        - staged-control system (km → cm → nm)
    - aspect sensing and control to 10's of micro-arcsec
    - positioning mirror surfaces to 5 nm
    - variable, non-condensing, continuous micro-Newton thrusters
- **precision metrology over multi-km baselines**
    - 2nm if used alone for pathlength control (no wavefront sensing)
    - 0.5 microns if hand-off to wavefront sensing & control for nm-level positioning
    - multiple modes to cover wide dynamic range
- **wavefront sensing and real-time, autonomous analysis**
- **methodologies for ground-based validation of distributed systems**
- **additional challenges (perceived as "easier" than the above)**
    - mass-production of "mirrorsat" spacecraft: cost-effective, high-volume fabrication, integration, & test
    - long mission lifetime requirement
    - light-weight UV quality mirrors with km-long radii of curvature (using active deformation of flats)
    - larger format (6 K x 6 K) energy resolving detectors with finer energy resolution (R=100)

Precision metrology over the long baselines required in interferometric missions like SI needs further development. Efforts are underway at JPL and SAO, but there is no assurance they will be supported as long as needed and to the fine levels required in the current long-term plan.

Wavefront sensing and control, based on feedback from the science data stream, especially in the context of a very sparse aperture imaging system, needs continued long-term work. The Fizeau Interferometer Testbed (FIT) is exploring this technology now with 7 elements and has plans to expand to as many as 18 elements, but it is a small effort that needs to be expanded to fully develop the needed algorithms and control laws. And it needs eventually to be integrated with a formation flying testbed, such as the FFTB (GSFC) or the SPHERES (MIT) experiment to develop and prove the staged-control laws needed to cover the full dynamic range from km's to m's to cm's to nm's.

Finally, one of the most challenging technology needs for SI and all large, distributed spacecraft missions: how does one test and validate on the ground, prior to flight a system whose components are numerous (~30) and whose separations are order of 100's of meters to many kilometers? This is also a critical need for Darwin, MAXIM (BHI), LF, and PI.

### 4.2 Development roadmap

The successful design and construction of the SI will rely on the development and validation of a number of critical technologies highlighted in the preceding sections. **Table 3** shows a high-level technology roadmap for these items.

Study of these technologies is ongoing at NASA/GSFC, JPL, SAO, various universities, and in industry, and significant leveraging and cross-fertilization will occur across projects, e.g. with JWST, Darwin/TPF, and LISA. A series of testbeds are in operation or are under development at GSFC, including the: Wavefront Control Testbed (WCT) to study image-based optical control methods for JWST, Phase Diverse Testbed (PDT) to study extended scene phase diversity optical control with moving array elements, Wide-Field Imaging Interferometry Testbed (WIIT) to study extending the field of Michelson imaging interferometers, and the Fizeau Interferometry Testbed (FIT) to study closed-loop control of an array of elements, as well as assess and refine technical requirements on hardware, control, and imaging algorithms. Studies of the full *SI* mission as well as Pathfinder concepts continue in GSFC's Integrated Design Center and Metrology Testbeds are under development at SAO (Ref. 2), JPL (Ref. 3), and GSFC (Ref. 4).

One of the more interesting technology options that is being pursued is an investigation of how much of the measurement and control job (of the various spacecraft and mirror surfaces in the distributed system) can be done purely by "external" (to the science data stream) metrology using, for example, lasers and at what point, and if, it will be necessary to handoff the measurement and control job to a system based on feedback from analysis of the science data stream. Our "baseline" mission concept in fact assumes that the external metrology system has measurement and command authority down to the millimeter or, if possible, the micron level and that a "closed-loop" optical control system, based on phase diversity analysis of the science data stream, takes over at smaller scales to obtain control down to the nanometer level. The exact point at which that handoff occurs in the multi-stage control system is one of the interesting points still to be resolved. Our technology development plan is based on pushing both technologies to their limits, i.e., driving the external metrology to the smallest attainable scales (effectively testing in the process if we can do the "entire job" this way) and driving the development of the wavefront sensing & control to the largest possible scales, in the hope that the two systems will in the end have a significant region of overlap in their control authority.

Table 3: Technology Roadmap for the Stellar Imager

| Technology Needed by SI | Development Plan and/or Candidate Technologies |
|---|---|
| Wavefront Sensing and Control | Phase Diverse Testbed (PDT), Fizeau Interferometry Testbed (FIT), Wavefront Control Testbed (WCT) |
| Closed-loop optical path control | Phase Diverse Testbed (PDT), Fizeau Interferometry Testbed (FIT) |
| Mass-production of spacecraft (SI "mirrorsats") | TBD (but see BATC approach in section 3.18 of full SI Vision Mission Report) |
| Lightweight, UV-quality mirrors with km-long radii of curvature | Chen (see full VM report), etc. |
| Large format energy-resolving UV detectors with resolution >100 | TBD – but driven by many missions |
| Methodologies for combining 20-30 simultaneous beams | Ground-based interferometers, FIT |
| Variable, non-condensing micro-newton thrusters | Field Emission Electric Propulsion units (FEEP's), etc. |
| Precision Formation Flying | GSFC Distributed Space Systems Roadmap (Figure 3.20 in full SI VM Report) |
| Aspect Control to 10's of micro-arcsecs | Trade external metrology vs. wavefront sen. |
| Precision Metrology over long baselines | JPL & SAO metrology labs |
| Methodologies/control processes for deployment and initial positioning of elements in large formations | GSFC Distributed Space Systems Roadmap (Figure 3.20 in full Report ) |

A space-based Pathfinder Mission would be extremely valuable in the path toward developing the full-scale SI. Existing useful precursor missions are limited: TPF-I, if it flies, will be a nulling, cryogenic interferometer operating in the infrared; SIM does not use the free-flying formations that will be needed for truly long-baseline facilities, and it will operate only at longer (optical) wavelengths. Furthermore, SIM will be used primarily as an astrometer, rather than as an imager. Formation flying issues may also be addressed by SMART3 and perhaps ST-9, though the content of these missions is still uncertain at this writing. It would therefore be desirable to have a Pathfinder mission with modest baselines (~20-50 m), a small number of primary elements (~3-5), decent size mirrors (~1 m), and the ability to perform ultraviolet beam combination and produce images in ultraviolet light. The small number of spacecraft/mirrors in this pathfinder mission would require extensive array reconfigurations and therefore limit observations to targets whose variability does not preclude long integrations. However, such a mission would both test most of the technologies needed for the full mission, as well as be capable of producing a significant scientific return. A pathfinder with 20-50 m baselines could, for example, image the surfaces of the apparently larger stars, such as the red supergiant Betelgeuse and many long-period variables (e.g. Mira), as well as symbiotic systems exhibiting mass-exchange between the components. The addition of high-resolution spectroscopy to such a mission could increase the science return even further at modest additional cost. One such Pathfinder mission design is described in the full SI Vision Mission Report, but the derivation of an optimal SI Pathfinder design will be the next step in our overall SI development process.

# 5. LAUNCH, DEPLOYMENT, AND OPERATIONS

## 5.1 Transportation to Sun-Earth L2 and deployment

SI can be launched using only one or two vehicles, depending on whether one or two hubs are included in the original deployment, and will be transferred to a Sun-Earth L2 libration orbit using a direct transfer trajectory. The coast phase (from Earth to L2) takes ~120 days. Mid-course correction maneuvers will be performed to correct any insertion energy errors and misalignments in the insertion orbit parameters. Upon arrival at the mission orbit, an insertion maneuver will be performed to balance the energy, allowing the spacecraft to be placed on the reference libration orbit. The size and orientation of the mission orbit for SI is not critical, therefore the maneuver (Delta-V) budget can be minimized for the mission lifetime. SI will be transferred to the mission orbit as one entity. Upon arrival and insertion into the mission orbit, the components will be deployed and maneuvered into their proper location, with careful collision avoidance procedures in place. This means that the relative navigation system and individual propulsion systems must be operating. The relative drift of the components will be in predictable directions, as the components will follow their own orbits and drift in patterns that are determined by the natural dynamics of the Sun-Earth libration region.

## 5.2 Operations

After initial check-out and commissioning, Stellar Imager will be autonomously controlled. Commands to re-point the system to a new target and reestablish the optical configuration at the end-point of the maneuver will come from a stored command area onboard. At each pointing, the onboard systems will automatically acquire guide stars, verify attitude, re-configure the arrays if needed, acquire the science target, and initialize the observing sequence. Data will be stored in onboard Solid State Recorders (SSRs) for later transmission to the ground. The Hub will contain the communications equipment for space-ground contact and be redundantly-designed for optimal lifetime. Optimally, there would be two Hubs in operation to ensure this "critical path" component has an immediately available backup. The availability of two Hubs also greatly increases the efficiency of the observatory – the second Hub can be pre-positioned while the first one is in use and the observatory can be re-pointed simply by tilting the primary array to align with the second Hub, without any large slews for the numerous (~30) mirrorsats. The SI design includes alternate communication capability for the unlikely event of a loss of primary Hub space-ground capabilities. SI will include onboard capability for recognizing failures in any given primary mirror unit and ability to avoid collision with the other units.

Communications will be through the Deep Space Network (DSN), and will be used to update onboard command memory, allow daily transmission of science and engineering data from the SSR(s) to the ground, collect tracking and ranging data for use in calculating orbital elements of SI, and send any re-configuration commanding. Primary uplinks to the Hubs will be at 2kbps using X band, with a 2kbps S band backup. The primary link will include automatic communications from the Hubs to each of the Mirrorsats using an SI internal communications subsystem. Backup link to the Mirrorsats will be via S band at 2kbps from the ground. All communication downlinks will also use the DSN. The nominal data rate from SI is about 125 Gb/day for ~11 months per year. This requires approximately one 30 minute Ka-band downlink per day. For ~1 month per year a data rate of about 250 Gb/day is expected, assuming a 2:1 lossless compression of the science data, which will require approximately one 60 minute Ka-band downlink per day. Primary downlink of stored data from the Hubs will be at 75 Mbps on Ka band. SI will automatically send data from the Mirrorsats to the Hub(s) for storage. Real-time data downlink from the Hubs will be via X band at 10 kbps with a backup of 6 kbps on S band. Backup real-time telemetry from the Mirrorsats directly to the ground will be via S band at 3 kbps.

# 6. CONCLUSIONS

The mission of the Stellar Imager is to enable an understanding of solar/stellar magnetic activity and its impact on the:

- origin and continued existence of life in the Universe
- structure and evolution of stars
- habitability of planets

and to study magnetic processes and their roles in the origin and evolution of structure and the transport of matter throughout the Universe. The SI Vision Mission Team has executed an ~1 year study to develop in detail the scientific goals and requirements of the mission, a baseline observatory architecture, the technology development needs of that and alternative architectures, a roadmap for that technology development, considered deployment and operations scenarios and addressed operations assurance and safety issues.

The study has shown that the scientific capabilities of such an ultra-high angular resolution UV/Optical interferometer are extraordinary, that credible design options are available, and that a sensible technology development path for supporting the development of the facility can be defined. SI fits well with the NASA and ESA strategic plans and complements other defined and conceptual missions, such as TPF, LF, and PI, and supports our collective desire as a species to understand extra-solar planetary systems and the habitability of surrounding planets, as well as improve our understanding of our own sun and its impact on earth's climate and it's future habitability.

Additional information on the Stellar Imager can be found at *http://hires.gsfc.nasa.gov/si/*

## 7. ACKNOWLEDGEMENTS

Additional contributions are gratefully acknowledged from the wide range of science and technology investigators and collaborators on the Stellar Imager Vision Mission Team (see **Table 4**). This work was supported, in part, by Vision Mission Study grants from NASA HQ to NASA-GSFC and from GSFC to Smithsonian Astrophysical Observatory, Seabrook Engineering, SUNY/Stonybrook, U. Colorado/Boulder, and STScI. Substantial complementary internal institutional support is gratefully acknowledged from all of the participating institutions.

Table 4: The Stellar Imager Vision Mission Team

- **Development led by NASA/GSFC in collaboration with:**

  | | |
  |---|---|
  | Ball Aerospace & Technologies Corp. | Lockheed Martin Advanced Tech. Center |
  | NASA's Jet Propulsion Laboratory | Naval Research Laboratory/NPOI |
  | Northrop-Grumman Space Technology | Seabrook Engineering |
  | Sigma Space Corporation | Smithsonian Astrophysical Observatory |
  | Space Telescope Science Institute | State Univ. of New York/Stonybrook |
  | Stanford University | University of Colorado at Boulder |
  | University of Maryland | University of Texas/Arlington |

- **Institutional and topical leads from these institutions include:**

  K. Carpenter (PI), C. Schrijver, R. Allen, A. Brown, D. Chenette, D. Mozurkewich, K. Hartman, M. Karovska, S. Kilston, J. Leitner, A. Liu, R. Lyon, J. Marzouk, R. Moe, N. Murphy, J. Phillips, F. Walter

- **Additional science and technical collaborators include:**

  T. Armstrong, T. Ayres, S. Baliunas, C. Bowers, G. Blackwood, J. Breckinridge, F. Bruhweiler, S. Cranmer, M. Cuntz, W. Danchi, A. Dupree, M. Elvis, N. Evans, C. Grady, F. Hadaegh, G. Harper, L. Hartman, R. Kimble, S. Korzennik, P. Liewer, R. Linfield, M. Lieber, J. Leitch, J. Linsky, M. Marengo, L. Mazzuca, J. Morse, L. Mundy, S. Neff, C. Noecker, R. Reinert, R. Reasenberg, D. Sasselov, S. Saar, J. Schou, P. Scherrer, M. Shao, W. Soon, G. Sonneborn, R. Stencel, B. Woodgate

- **International Collaborators include:**

  J. Christensen-Dalsgaard, F. Favata, K. Strassmeier, O. Von der Luehe

- **Student Participants include:**

  Linda Watson (undergrad-Univ. Florida/CfA), Darin Ragozzine (undergrad-Harvard, grad-CalTech), Mikhail Dhruv (high school), Fonda Day (undergrad/CU)